\documentclass[10pt]{article}
\usepackage[dvips]{graphicx,color}
\usepackage{amsmath,amssymb}
\usepackage{yfonts}
\newcommand{\bce}{\begin{center}}
\newcommand{\ece}{\end{center}}
\newcommand{\be}{\begin{equation}}
\newcommand{\ee}{\end{equation}}
\newcommand{\bea}{\begin{eqnarray}}
\newcommand{\eea}{\end{eqnarray}}
\newcommand{\bdes}{\begin{description}}
\newcommand{\edes}{\end{description}}
\newcommand{\bit}{\begin{itemize}}
\newcommand{\eit}{\end{itemize}}
\newcommand{\btt}{\begin{tt}}
\renewcommand{\baselinestretch}{1.2}
\renewcommand{\thesection}{\arabic{section}.}

\renewcommand{\theequation}{\arabic{equation}}

\newcommand{\E}{\> = \>}
\newcommand{\EA}{&=&}
\newcommand{\non}{\nonumber \\}

\newcommand{\sgn}{{\rm sgn}}
\newcommand{\Eqi}{\> \equiv \>}
\def\Uint{\int_{-\infty}^{+\infty}}
\def\Def{\> := \>}
\def\deF{\> =: \>}
\def\la{\left\langle \,}
\def\ra{\, \right\rangle}
\def\lrp{\left ( \, }    
\def\rrp{\, \right ) }   
\def\lsp{\left [ \, }    
\def\rsp{\, \right ] }   
\def\lcp{\left \{ \, }   
\def\rcp{\, \right \} }  
\def\lvl{\, \left | \, }     
\def\rvl{\, \right | \, }    
\renewcommand{\theequation}{\thesection\arabic{equation}}

\newcounter{abb}

\begin{document}
\begin{flushright}
PSI-PR-21-07\\
\hspace{0.0cm}\\
\end{flushright}
\setcounter{page}{1}
\thispagestyle{empty}

\vspace{2cm}

\bce
{\Large\bf On the numerical evaluation of real-time path integrals:}\\

\vspace{0.3cm}

{\large\bf Double exponential integration and the Maslov correction}\\

\vspace{2cm}

{\large R.~Rosenfelder}\\

\vspace{2cm}

\noindent
Particle Theory Group, Paul Scherrer Institute,
CH-5232 Villigen PSI, Switzerland\\

\ece

\vspace{1cm}

\begin{abstract}
\noindent
Ooura's double exponential integration formula for Fourier transforms is applied to the 
oscillatory integrals occuring in the path-integral  description of real-time Quantum Mechanics.
Due to an inherent, implicit regularization 
multi-dimensional Gauss-Fresnel integrals are obtained numerically with high precision
but modest number of function calls. In addition, the Maslov correction for the harmonic oscillator is evaluated numerically with an increasing number of time slices in the  path integral thereby clearly demonstrating that the real-time propagator acquires an additional phase $ - \pi/2 $ each time the particle passes through a focal point. However, in the vicinity of these singularities an overall small damping factor is required. Prospects of evaluating scattering amplitudes of finite-range potentials
by direct numerical evaluation of a real-time path integral are discussed.

\end{abstract}
\vspace{3cm}

\newpage


\section{Introduction}
\setcounter{equation}{0}

Among the many formulations of Quantum Mechanics the path-integral method 
(see e.g. \cite{Feyn}, \cite{Klein}, \cite{engpath}) is a very attractive one as 
it gives rise to new insights, approximation schemes and
can be readily generalized to Many-Body Physics and Quantum Field Theory. However, 
for numerical evaluation 
a path (i.e. functional, i.e. infinite-dimensional) integral poses an extraordinary challenge requiring at present (unphysical) imaginary time and Monte-Carlo methods. Still, due to the heavy oscillations of the integrand in real time scattering cannot be treated in this way (apart from low-energy quantities) and one has to resort to
semi-classical approximative methods (see, e.g. Ref. \cite{Makri}).

In a broader context "numerical integration" is essential in all quantitative sciences, 
not only in theoretical and computational physics. It is "a wide field" \cite{Hahn} with a vast literature which cannot be cited adequatly here (a standard textbook is Ref.  \cite{Rabin})
and many different methods. 

It is somehow surprising that after centuries of work on numerical integration rules
(associated with the names of Kepler, Newton, Simpson, Lagrange, Gauss and others)
a new contender for the "best" all-purpose method appeared:  
the double exponential (DE) integration (or tanh-sinh quadrature) method of Takahasi and Mori \cite{TaMori} (see also Ref. \cite{Mori-discov}). This method has become
a standard tool to obtain high-precision results with only $ \> {\cal O}(N \ln  N) \> $ function calls \cite{BBBZ}.  This is due to 
a transformation $ x = \phi(t) $ which maps a finite interval into an infinite one 
and leads to a new integrand which decays asymptotically with double exponential rate so that 
the integral can be best approximated by the extended trapezoidal rule.
Although other integration schemes (like Gaussian quadrature) are superior for smooth integrands, it can be argued that the DE-method comes close to a all-purpose quadrature scheme \footnote{A caveat: in certain applications the method fails to give
accurate results \cite{num osc}. This is not surprising as realistically one may doubt whether  
a "best" scheme for {\it all} functions exists 
.}.


Ooura and Mori \cite{OoMori} have extended this scheme to oscillating integrands and found spectacular results.
For example, Euler's constant is obtained from integrating
\be
\int_0^{\infty} dx \> \lrp - \ln x \rrp \, \sin x \E \gamma_E
\label{gamma_E}
\ee
to an absolute accuracy  of $ 10^{-12} $ with only $80$ function calls 
(see Table 2 in Ref. \cite{OoMori})
despite the fact that the integral only exists as limiting case
\be
\lim_{\eta \to 0}  \int_0^{\infty} dx \> \lrp - \ln x \rrp \, \sin x \, e^{-\eta x} \> .
\ee
Thus the double exponential method implicitly introduces a suitable regularization scheme. 
This makes it a very 
appealing quadrature rule for quantum-mechanical path integrals in real time 
\be
\int {\cal D} x(t) \> \exp \lcp i \int_{t_a}^{t_b} dt \lsp \frac{m}{2} \dot x^2(t) - V(x(t)) \rsp \rcp
\ee
where the convergence of these 
(Fresnel-type) integrals is ensured by Feynman's ``$ i 0^+ $-rule" 
which either modifies the mass 
$ m \to m + i 0^+ $ or the squared frequency in a harmonic oscillator potential 
$ \> V = m \Omega^2 x^2/2 \> $ 
as $ \Omega^2 \to \Omega^2 - i 0^+ $. As a free field theory can be seen as a system of coupled oscillators 
it is not surprising that the latter prescription (here for the squared mass) is needed to specify 
the singularities of Green functions. 
\vspace{0.1cm}

It is the purpose of this work to show that a direct evaluation of 
multi-dimensional oscillating integrals is possible by applying the double exponential
methods alluded to before. This will be demonstrated by calculating high-dimensional Gauss-Fresnel integrals and
by evaluating the so-called Maslov phase for a non-relativistic particle in a harmonic oscillator 
well where the exact solutions are readily avalable.

\section{Double exponential integration}
\setcounter{equation}{0}
Ooura and Mori use a transformation $ \phi(t) $ such that
the derivative $ \phi'(t)$ goes to zero double exponentially for $ t \to -\infty $ while the function 
 $ \phi(t) $ approaches $ t $ double exponentially  as $ t \to +\infty $. The latter property ensures that
for large $ t $ the zeroes of the oscillating integrand are nearly hit leading to the fast convergence of the scheme.

Ooura has later given an improved version of this method \footnote{See Fig. 3 in Ref.  \cite{OMrobust} for a comparison
with the old method for the integral in Eq. \eqref{gamma_E}.}
 which I will use in the following, viz.
"Approximation formula 2" (Eq. (3.5) in Ref.\cite{Ooura})
with the simplifications $ \omega_0 = \omega $ and $ N_+ = N_- \Eqi k_{\rm max} $. Although originally only given for 
$ \> \omega > 0 \> $ one  can extend it also to the case $ \omega < 0 $ by complex conjugation provided the function 
$ \> f(x) \> $ is real. Thus
\be 
\boxed{
\int_0^{\infty} dx \> f(x) \, e^{i \omega x} \> \simeq \> 
\frac{\pi}{|\omega|} \> \> \sum_{k = - k_{\rm max}}^{k = + k_{\rm max}} \! f \lrp \frac{\pi}{|\omega| h} \phi(k h ) 
\rrp \> \phi'( k h ) \,  \Big \{ \exp \lsp i \, {\rm sgn}(\omega)\, \frac{\pi}{h} \, 
\phi(k h)\rsp - (-1)^k \Big \} 
\label{Ooura form2}
}
\ee
where  $ \> {\rm sgn}(\omega) \Def \omega/|\omega| \> $ and
\be 
\phi(t) \E \frac{t}{1 - \exp\lsp -2 t - \alpha \lrp 1 - e^{-t} \rrp - \beta \lrp e^t - 1 \rrp \rsp } \> .
\ee
Ooura's parameter $ \alpha, \beta $ are $ \omega-$dependend:
\be 
\beta \E \frac{1}{4} \> , \qquad 0 \>  < \> \alpha \E \beta \> \sqrt{\frac{4 \,|\omega| h}{4 \, |\omega| h  + 
\ln \lrp 1 + \frac{\pi}{|\omega| h} \rrp}} \quad < \quad \beta \> .
\label{alpha beta}
\ee
Note that the case $ \> \omega = 0 \> $ is undefined in Ooura's integration rule although
one would then expect that it reduces to the standard double exponential integration rule
for an half-infinite interval. However, 
questions about the allowed frequency range or the class of admissible functions are outside the scope of the present work.

\vspace{0.2cm}

\section{Application I: Multi-dimensional Gauss-Fresnel integral}
\setcounter{equation}{0}
I define the simplest $N$-dimensional Gauss-Fresnel integral as
\be 
G\!F_N(\omega) \Def \int d^N y \> \exp \lrp i \omega \sum_{k=1}^N y_k^2 \rrp \> , \quad \omega \quad {\rm real} \> .
\ee
Its analytical value is obtained by a regularization 
\be 
G\!F_N(\omega) \E \lim_{\eta \to 0} \, \prod_{k=1}^N \lrp \int d^N y \> \exp \lsp i \lrp \omega + i \eta \rrp  y^2 \rsp \rrp\E \lim_{\eta \to 0} \lrp \frac{\pi i}{\omega + i \eta} \rrp^{N/2} \E \lrp \frac{\pi}{|\omega|} \rrp^{N/2} \, \exp \lrp i  \, {\rm sgn}(\omega) \,\pi \frac{N}{4}\rrp
\label{GF}
\ee
where the positive sign of the square root has to be taken (for a thorough mathematical treatment see, e.g. Refs. \cite{NagaMiya1}, \cite{NagaMiya2}).
\vspace{0.2cm}

I will calculate the Gauss-Fresnel integral in {\it hyperspherical co-ordinates}. As the integrand only depends on 
the hyperradius $ \> R = (\sum_{k=1}^N y_k^2 )^{1/2} \> $ one obtains
\be 
G\!F_N(\omega) \E S_{N-1} \cdot \int_0^{\infty} dR \, R^{N-1} \, \exp \lrp i \omega R^2 \rrp
\ee
where
\be 
S_{N-1} \E \frac{2 \pi^{N/2}}{\Gamma (N/2)}
\label{S_(N-1)}
\ee
is the $(N-1)$-dimensional surface of the $N$-sphere. The variable change $ R = \sqrt{y} $ brings the integral into the form
\be 
G\!F_N(\omega) \E  \frac{1}{2} S_{N-1} \int_0^{\infty} dy \> y^{N/2-1} \> e^{i \omega y}
\ee
to which Ooura's numerical integration formula \eqref{Ooura form2} will be applied. Note that
this application is much more demanding than the numerical evaluation of Euler's number by  Eq. \eqref{gamma_E} as now the integrand grows power-like instead of logarithmically.

\vspace{0.2cm}

\begin{figure}[htbp]
\bce
\includegraphics[angle=0,scale=0.5]{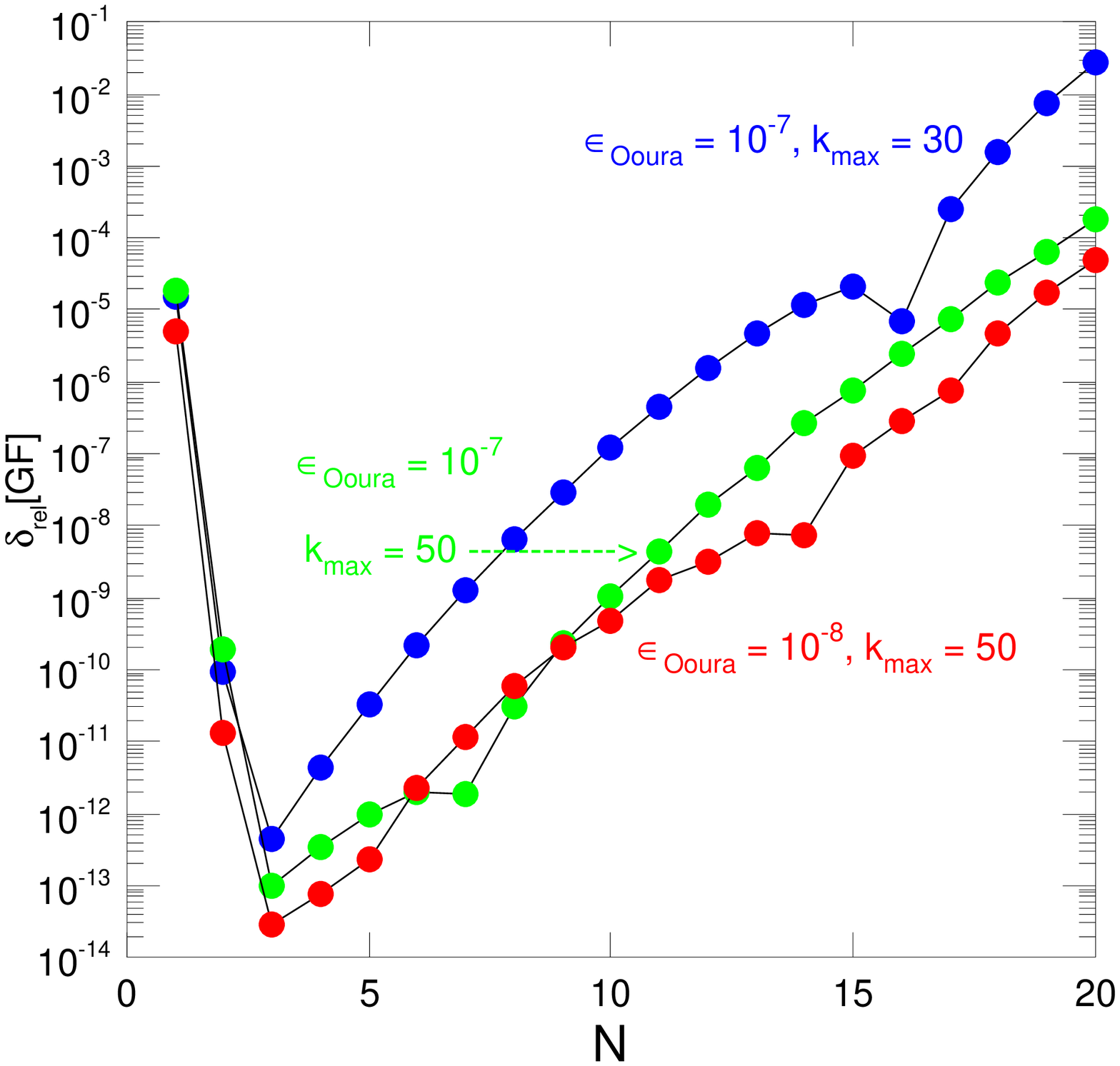}
\label{abb:1}
\ece
\vspace*{-0.5cm}

{\bf Fig. \arabic{abb}} : {\small Relative complex deviation (as defined in Eq. \eqref{rel complex dev}) of the Gauss-Fresnel integral evaluated numerically with Ooura's method
from the exact value for $ \omega = \pm 1 $ and increasing dimension $N$. 
Results for different accuracy parameters (see Table 1)  are depicted.}
\end{figure}

Fig. \ref{abb:1} shows how much the numerical result (for $ \> \omega = \pm 1 \> $ ) deviates from the exact one. Here and in the following I use as metric the "relative complex deviation"
\be 
\delta_{\rm rel}[G\!F] \Def \lvl \frac{G\!F_{\rm num} - G\!F_{\rm exact}}{G\!F_{\rm exact}} \rvl
\label{rel complex dev}
\ee
\vspace{0.2cm}

\refstepcounter{abb}

\vspace{1cm}

\noindent
which quantifies the agreement of the numerical calculation compared with the exact result both 
for the absolute magnitude as for the phase of a complex quantity. Obviously  a very good agreement
with the exact values is obtained in a wide range of dimensions  -- without any regularization!

\section{Application II: Maslov correction for the harmonic oscillator}
\setcounter{equation}{0}

It is long known \cite{Mas} and covered in many textbooks  (e.g. in Ref. \cite{Schul}, ch. 17 or Ref. \cite{Klein}, p. 100) but frequently overseen \footnote{In particular in some quantum field theory texts where the harmonic oscillator serves as an example of exact path integration, e.g. in Refs. \cite{Das}, \cite{Wen}, \cite{BTov} or in the prepublication version of Ref. \cite{Woit}. A detailed exposition of the Maslov correction is given in Ref. \cite{Horv}.
An early proof for this phenomenon was provided by Pechukas \cite{Pech} who evaluated  the time-evolution operator at time $ \pi < \Omega T < 2 \pi $ with the help of the composition law 
$ \> \hat U(T,0) = \hat U(T,T') \, \hat U(T',0) \> $  with $ 0 < \Omega (T-T'), \Omega T' < \pi \> $
(see also Problem 9 in Ref. \cite{engpath}).
} that the matrix element of the time-evolution operator for a non-relativistic particle in a harmonic potential 
\be
V^{\rm h.o.} (x) \E  \frac{m}{2} \Omega^2 x^2 \> 
\ee
as given by
\be 
\la x_b \lvl \hat U^{\rm h.o.}(t_b,t_a) \rvl x_a \ra \E F^{\rm h.o.}(T = t_b - t_a) \, \cdot \, \exp \lsp i S_{\rm class}^{\rm h.o.}(x_b,x_a;T) \rsp  
\label{U h.o.}
\ee
acquires an 
additional phase each time the particle passes through a {\it focal point} at $ \> \Omega T = n \pi \> $ with $ \> n = 1,2 \ldots $.
In Eq. \eqref{U h.o.} $  S_{\rm class}^{\rm h.o.}(x_b,x_a;T) $ is the classical action of the harmonic oscillator (for an explicit expression see, e.g. eq. (1.64) in Ref. \cite{engpath}) and 
\be
 F^{\rm h.o.}(T) \E \lrp \frac{m \, \Omega}{2 \pi i \lvl \sin(\Omega T) \rvl} \rrp^{d/2} \> e^{- \frac{1}{2} i \pi \, d \, n_{\rm M}} \> , \quad
 n_{\rm M} \E \sum_{n=1}^{\infty} \Theta ( \Omega T - n \pi ) 
 \label{F exact}
\ee
the prefactor in $ d $ space dimensions which is due to the quantum fluctuations ( $ \Theta(x) $ is the step or Heaviside function and a system of units is used in which $ \> \hbar = 1 \> $).
Obviously, the prefactor diverges at
these focal points (creating the so-called caustics)
and the particle starts anew its quantum-mechanical propagation (like "a phoenix rising from the ashes") 
but with an additional phase $ - d \pi/2 $ as sole remainder of its previous history. 

\vspace{0.1cm}
 
It should be emphasized that the Maslov correction is a genuine quantum-mechanical phenomenon occuring only in real time: The euclidean
version (for example the partition function) does not display it \footnote{It is unclear to me 
how an analytic continuation from euclidean to real time will generate the Maslov phase. According to the 
Osterwalder-Schrader reconstruction theorem \cite{OSchr}, extensively used in Quantum Field Theory for justifying Wick 
rotations, this should be possible.}. 
The occurence of the Maslov correction is easily seen in the Fourier path integral for the harmonic oscillator propagator, see e.g. ch. 1.2 in Ref. \cite{engpath}. 

In the mathematical literature this is a well-known result: Nagano and Miyazaki \cite{NagaMiya1}
refer to several textbooks and
cite an article by H\"ormander from 1971 as earliest reference \footnote{Thus calling the
correction the "Maslov phase" seems to be not fully correct but only underlines the saying:
{\it Most named effects in physics are not named after the first discoverer...}}. In particular, their Proposition 2.5  (for the particular case $\xi = 0 $) 
\be 
\frac{1}{(2 \pi)^{N/2}} \int d^Nx \, \exp \lsp \frac{i}{2} \sum_{j=1}^N x_j \, {\cal A}_{j k} \, x_k 
\rsp
\E \frac{\exp \lrp i \frac{\pi}{4} \, {\rm sgn} {\cal A} \rrp}{|\det {\cal A}|^{1/2}}
\label{Hoermander}
\ee
is the generalization of Eq. \eqref{GF}. Here
\be 
{\rm sgn}\,  {\cal A} \E n_+ - n_-
\ee
denotes the {\it signature} (Ref. \cite{HoJo}, p. 221) of the real, symmetric, non-singular matrix $ {\cal A} $, i.e. 
the difference of positive and negative eigenvalues. Using $ n_+ + n_- = N $ the phase factor in Eq. \eqref{Hoermander} thus is
\be 
\exp \lrp i \frac{\pi}{4} N \rrp \, \cdot \, \exp \lrp -i \frac{\pi}{2} n_- \rrp \> .
\ee
While the first factor also shows up in the free propagator
the second factor obviously describes the effect of negative eigenvalues, i. e. the Maslov phase.

\subsection{The Maslov correction in the time-sliced path integral}

%
%
I will be using the usual time-slicing method for evaluating the path integral 
in $ d = 1 $ dimensions (see, e.g. Ref.\cite{engpath}, section I.2) but with $ N + 1 $ intervals
to simplify the notation. For accelerating the convergence of the discretized version to the continuum one, I will utilize the symmetric form of the Trotter product formula eq. (1.33) 
\be
\exp \left [ -i \Delta T ( \hat T  + \hat V) \right ]
\> = \exp \left [ -i \Delta T \hat V / 2 \right ] \> \cdot \>
\exp \left [-i \Delta T \hat T  \right ] \> \cdot \>
\exp \left [ -i \Delta T \hat V / 2 \right ] \> + {\cal O}((\Delta T)^3)
\> ,
\label{kurz Zeit sym}
\ee
applied to kinetic ($\hat T$) and potential ($\hat V$) energy operator.
This is correct up to order
$ (\Delta T)^2 $ where $\Delta T = (t_b - t_a)/(N+1) $ is the time-step of the discretized version \footnote{Expressions which are correct to arbitrary high orders have been investigated in 
Refs. \cite{Serbia 1}, \cite{Serbia 2} in terms of increasingly higher derivatives of the potential. In lattice field theories this amounts to constructing "improved actions", an example of these is given in Ref. \cite{BorRo}.}. 
This slightly modifies eq. (1.28) in that reference
to
\bea
\hspace{-1cm} U(x_b, t_b; x_a, t_a) \EA \lim_{N \to \infty} \>
\left (\frac{m}{2 \pi i \Delta T} \right )^{(N+1)/2}
\int_{-\infty}^{+\infty} dx_1 \> dx_2 \> ... \> dx_N \non
&& \cdot \exp \left \{ i \Delta T \sum_{j=1}^{N+1} \left [
\frac{m}{2} \left ( \frac{x_{j} - x_{j-1}}{\Delta T} \right )^2 - \frac{1}{2}
V(x_j) - \frac{1}{2} V(x_{j-1}) \right ] \> \right \} 
\label{Lagrange exterior} 
\eea
where $ x_0 = x_a $ and $ x_{N+1} = x_b $ are fixed. Eq. \eqref{Lagrange exterior} thus is based on  an "exterior" average $ \> \frac{ V(x_j) + V(x_{j-1})}{2} \> $ for the potential whereas the ususal "midpoint rule" with
$ \> V \lrp \frac{x_j + x_{j-1}}{2} \rrp \> $  is an "interior" average. In the continuum limit 
$ \> \Delta T \to 0 , N \to \infty \> $ there is, of course, no difference between these distinct discretization, but for finite $ \Delta T $ there is. For simplicity, in the present note, I will use Eq. \eqref{Lagrange exterior}.
\vspace{0.2cm}
 
The prefactor $ F(T) $ for the harmonic oscillator potential is thus obtained by putting 
$ x_a = x_b = 0 $ so that the classical action vanishes as well as terms with 
$ V^{\rm h. o.}(0) = 0 $. 
For $ N \ge 1 $ this gives 
\be 
F^{\rm h. o.} (T) \E U(0, T; 0, 0) \E \lim_{N \to \infty} \, F_N^{\rm h.o.}(T)
\ee
with 
\bea
 F_N^{\rm h.o.}(T) &=& \left (\frac{m}{2 \pi i \Delta T } \right )^{\frac{N+1}{2}}
\int_{-\infty}^{+\infty} \! \! dx_1 \> dx_2 ... dx_N \,
 \exp \Bigg \{ \frac{i m}{2 \Delta T} \Bigg [ x_N^2 + x_1^2
+ \sum_{j=2}^N  \big ((x_j - x_{j-1} \big)^2 - (\Omega \Delta T)^2 \sum_{j=1}^N x_j^2 \Bigg ] 
 \Bigg \} \non
\EA  \left (\frac{m}{2 \pi i \Delta T } \right )^{(N+1)/2}
\int_{-\infty}^{+\infty} dx_1 \> dx_2 \> ... \> dx_N 
\> \exp \left \{ \frac{i m}{\Delta T} \left [ \xi \sum_{j=1}^N x_j^2 - 
\sum_{j=2}^N  x_j \cdot  x_{j-1} \right ] \> \right \} 
\label{pref F_N}
\eea
where I have defined 
\be 
\xi \Def 1 - \frac{1}{2} (\Delta T)^2 \Omega^2 \E 1 - \frac{1}{2} \, \frac{\Omega^2 T^2}{(N+1)^2}
\label{def xi}
\ee 
and an empty sum is to be taken as zero.
%
%

Since these are all Gauss-Fresnel integrals they can be evaluated analytically also for finite 
$ N $ as demonstrated in Appendix A.
Let us write the prefactor as in eq. (1.89) of Ref. \cite{engpath}
\be 
F_N^{\rm h.o.} \deF \lrp \frac{m}{2 \pi i} \frac{1}{|f_N (\Omega T)|} \rrp^{1/2} \, 
e^{i \Phi_N^{\rm Maslov}(\Omega T)} \> .
\label{prefactor square root}
\ee
Then one finds for finite $ N $ 
\be 
\Omega \, f_N(\tau) \E \frac{\tau}{N+1} \, U_N \lrp 1 - \frac{\tau^2}{4 (N+1)^2} \rrp
\> , \quad \Phi_N^{\rm Maslov}(\tau) \E - \frac{\pi}{2} \sum_{n=1}^N \Theta \lsp \tau - 2 (N+1) \sin \lrp
\frac{n \pi}{2 (N+1)} \rrp \rsp
\label{f_N}
\ee
where 
\be 
\tau \Def \Omega T 
\label{def tau}
\ee 
is the dimensionless time and $ U_N(z) $ the Chebyshev polynomial of the second kind. In Appendix A the continuum limit ( $ N \to \infty $, $ T $ fixed ) 
\be 
\Omega \, f_{\infty}(\tau) \E \sin \tau \> , \quad \Phi_{\infty}^{\rm Maslov}(\tau) \E - \frac{\pi}{2} \, 
\sum_{n=1}^{\infty} \Theta \lrp \tau - n \pi \rrp 
\label{exact no damp}
\ee
is also studied in detail.
\vspace{0.5cm}

\subsection{Numerical implementation and results}

The challenge is now to evaluate Eq. \eqref{pref F_N} with a finite number of time slices, i.e. a finite 
number $ N $ of intermediate integrals. To cope with the oscillating integrand I use
$N$-dimensional spherical coordinates
\bea 
x_1 \EA R \, \cos ( \phi_1 ) \non
x_2 \EA R \, \sin ( \phi_1 ) \, \cos ( \phi_2 ) \non
x_3 \EA  R \, \sin ( \phi_1 ) \, \sin ( \phi_2 ) \, 
\cos ( \phi_3 )  \non
\vdots \non
x_{N-1} \EA  R \, \sin ( \phi_1) \ldots \sin ( \phi_{N-2} ) \, 
\cos ( \phi_{N-1} ) \non
x_N \EA  R \, \sin ( \phi_1 ) \ldots \sin ( \phi_{N-2} ) \, 
\sin ( \phi_{N-1} )
\eea 
so that the most violent oscillations come from the infinite integral over the hyperradius as in
the  Gauss-Fresnel integral. Whereas $ \> R \in [0,\infty] \> $, 
the integration over the angles $ \phi_j $  is restricted:
\be 
\phi_1, \phi_2 \ldots \phi_{N-2} \in [0,\pi] \quad  \mbox{but} \quad 
\phi_{N-1} \in [0, 2 \pi ] \> , 
\ee
The volume element is
\be 
d^Nx \E R^{N-1} \, \sin^{N-2} ( \phi_1 ) \, \sin^{N-3} ( \phi_2 ) 
\ldots
\sin ( \phi_{N-2} )  \> dR \, d\phi_1 \, d\phi_2 \ldots d\phi_{N-1} \E R^{N-1} dR \> d\Omega_{N-1}
\> .
\ee
Substituting  $ \> R = \sqrt{y \Delta T/m} \> \> $ 
Eq. \eqref{pref F_N} then reads
\be 
F_N^{\rm h.o.}(\tau) \E \sqrt{\frac{m}{8 \pi i \Delta T} } \, (2 \pi i)^{-N/2}
\, \int d\Omega_{N-1} \, \int_0^{\infty} dy \, y^{N/2 - 1} \>  
\exp \lsp i \, \omega_N \lrp \phi_1 \ldots \phi_{N-1};\tau \rrp \, y \rsp
\label{FN hyper}
\ee
where
\be 
\omega_N \lrp \phi_1 \ldots \phi_{N-1};\tau \rrp  \E \lrp \xi \sum_{j=1}^N x_j^2 - \sum_{j=2}^N
x_j \cdot x_{j-1} \rrp \Bigg / R^2 \E \xi  - \sum_{j=2}^N
x_j \cdot x_{j-1} \Big / R^2 \>.
\label{omega_N}
\ee
By normalizing the prefactor to the free case trivial complex factors
are eliminated. The free case is easily obtained by letting $ \> \Omega \to 0 \> $ in Eq. \eqref{f_N} 
so that
$ f_N \>  \to  T/(N+1) \> , \Phi_N^{\rm Maslov} \> \to  0 \> $. Thus
\bea
F_N^{\rm h.o.}(\tau) \big / F_N^{\rm free} \EA \frac{1}{2} \, (2 \pi i)^{-N/2}
\, \int d\Omega_{N-1} \, \int_0^{\infty} dy \, y^{N/2 - 1} \>  
\exp \lsp i \, \omega_N \lrp \phi_1 \ldots \phi_{N-1};\tau \rrp \, y \rsp \non
\EA \sqrt{\frac{\Delta T}{|f_N(\tau)|}} \, e^{i \phi_N^{\rm Maslov}(\tau)}
\label{FN hyper 2}
\eea
directly gives magnitude and phase of the prefactor. Note that the phase is determined only 
modulo $ 2 \pi $, or {\it vice versa}
\be
\phi_N^{\rm Maslov} \E {\rm atan}2\lrp {\rm Re} F_N,{\rm Im} F_N \rrp + 2  \pi  \, n \quad, \qquad
n \E 0, \pm 1, \pm 2 \ldots
\label{Phi ambiguity}
\ee
(see Eq. \eqref{atan2}). We are free to "align", i. e. choose $ n $ in order to directly compare with the analytic result \eqref{exact no damp} without changing the physics.
\vspace{0.2cm}

The strategy to evaluate Eq. \eqref{FN hyper 2} numerically is the following:
Perform the integration over the 
hyper-radius $ \> R \> $ by means of Ooura' s integration formula for oscillatory integrals while
the integration over the angles is done by standard integration routines. This is summarized
in Table 1 which also lists the relevant accuracy parameters in these routines and their typical values.
%
%

\vspace{1cm}

\renewcommand{\baselinestretch}{0.9}
\small\normalsize
\bce
\begin{table}[htb]

\small
\begin{tabular}{l|l|c|rl|c} \hline
                      &                  &            &                 &             
                      &                 \\

 routine/method       & application/feature& Ref.     &  accuracy       & \qquad explanation   
                      &  typical         \\ 
                      &                  &            &  parameters     &                
                      &  value           \\
                      &                  &            &                 &               
                      &                  \\ \hline
                      &                  &            &                 &              
                      &                  \\

{\small\bf Ooura}     & double exponential &  \cite{Ooura} &  $k_{\max} \>  $:   &  summation cut-off 
                      &   50 -- 80         \\
                      & for oscill. integrand &           &               & in Eq. (2.1)
                      &                   \\
                      & deterministic    &    & $\epsilon_{\rm Ooura} \>  $: & smallest weight (Eq.(B.6))            & $ 10^{-8}$      \\
					  &                 &             &                 &             
					  &                 \\\hline
					  &                 &             &                 &        &  \\
{\small\bf Double Exponential} & general integrand &  \cite{TaMori}   & $k_{\rm max} \>  $:  & as in {\bf Ooura} 
                      & $ 50 $          \\
                      & deterministic   &      & $\epsilon_{\rm DE} \>  $:        & smallest weight
                      &   $ 10^{-8} $    \\ 
					  &                 &             &                 &             
					  &                 \\
{\small\bf Gauss-Chebyshev}& adaptive&  \cite{PSM} & $m_{Cheby} \>  $: & max. \# of function calls
                      &  $ 511 $  \\  
 (modified for complex &  deterministic  &       & $\epsilon_{\rm Cheby} \>  $: & required
  rel. accuracy       & $10^{-3}$ \\ 
  integrand)           &               &               &                 &             & \\
					  &               &               &                 &        
					  &               \\
{\small\bf VEGAS}     & importance sampling &  \cite{Vegas}     & $n_{\rm call} \>  $:  & max. \# of function calls
                      &   $10^5 $ -- $10^6$  \\              
                      &  Monte-Carlo  &          &  $itmx \>  $:        & max. \# of iterations 
                      &          $10$ \\ 
                      &               &               &                 & 
                      &               \\ \hline
\end{tabular}

\vspace*{1cm}
\small\normalsize
\normalsize
{\bf Table 1}: Integration routines used (in order of appearance) with typical values of the corresponding accuracy \hspace*{1.5cm} parameters.

\label{table:1}
\end{table}
\ece

\renewcommand{\baselinestretch}{1.2}

\refstepcounter{abb}

\begin{figure}[hbt]
\vspace*{-1cm}
\bce
\includegraphics[angle=0,scale=0.5]{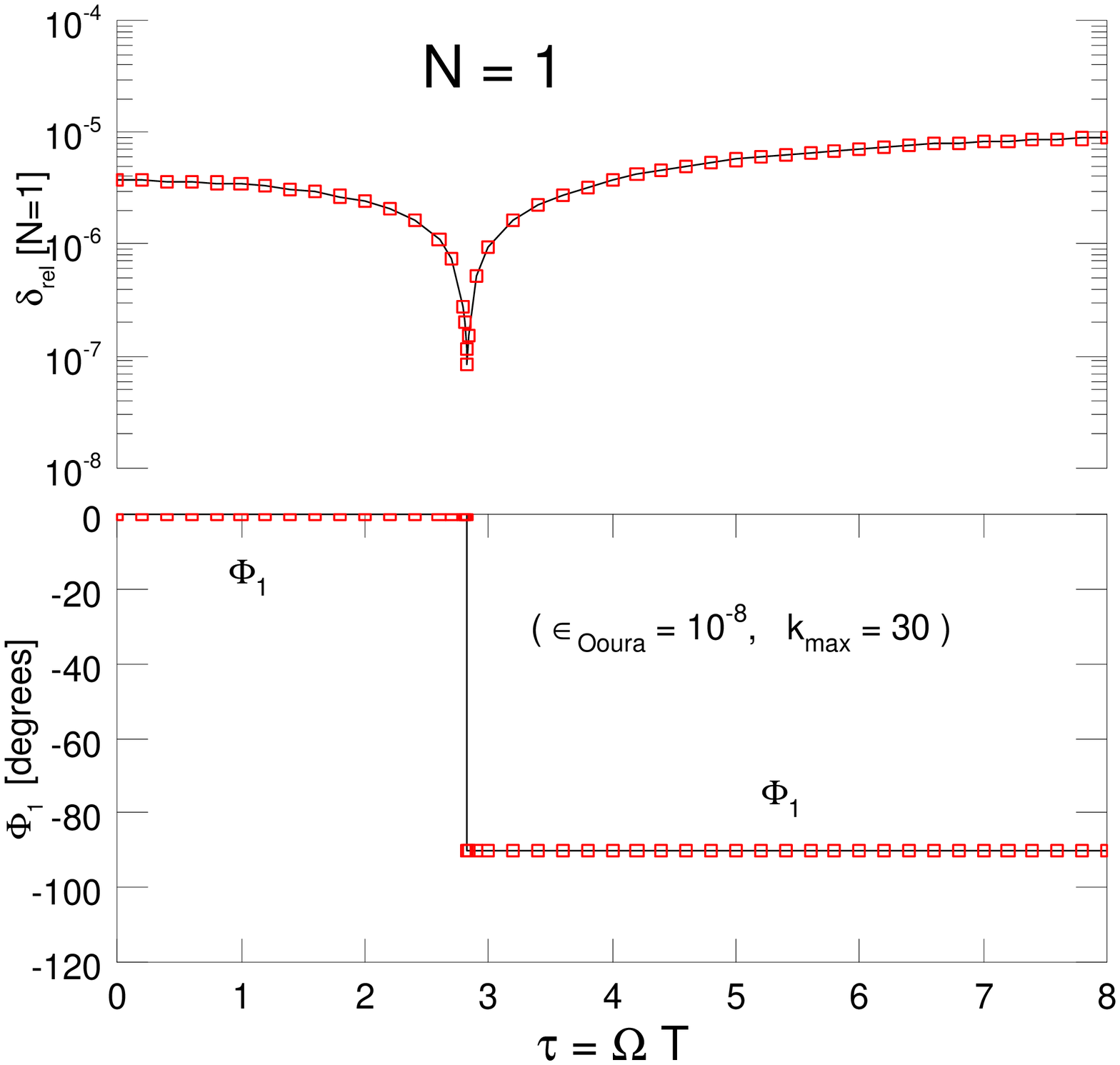}
\label{abb:2}
\ece

{\bf Fig. \arabic{abb}} : {\small Upper panel: Relative complex deviation (as defined in Eq. 
\eqref{rel complex dev})
of the numerical result for the prefactor of the harmonic oscillator propagator with $ \> N = 1 \> $
(i.e. 2 time slices in the path integral) from the exact $(N = 1)$-result. In the lower panel the resulting additional {\it Maslov} phase $ \> \Phi_1 \> $
is plotted as 
as function of the dimensionless time
$ \tau = \Omega T $ where $\> \Omega \> $ denotes the oscillator frequency. 
The integral 
\eqref{FN hyper 2} was evaluated numerically with Oura's formula \eqref{Ooura form2} and is compared with the exact $(N=1)$-result (black line) which has
one focal point at $ \> \tau = \sqrt{8} = 2.818... \> $.}
\end{figure}

\vspace{0.2cm}

\noindent
\normalsize
Let us start discussing the numerical results with the simplest (but already non-trivial) case 
$ \> N = 1 \>$, i.e. 2 time slices for the discretized path integral and thus no integral over
an hyperspherical angle.
Fig. \ref{abb:2} shows the numerical result when evaluating Eq. \eqref{FN hyper} for 
$ \> N = 1 \> $, i.e.
\be 
F_1^{\rm h.o.}(\tau) \E \sqrt{\frac{m}{2 \Delta T}} \, \frac{S_0}{2 \pi i } \, \int_0^{\infty} dy \, 
y^{-1/2} e^{i \omega_1 y } \E \sqrt{\frac{m}{2 \pi i \Delta T}} \, \frac{1}{\sqrt{\xi + i 0^+}}
\ee
since $ \> S_0 = 2 \> $ (see Eq. \eqref{S_(N-1)}) and $ \> \omega_1 \equiv \xi = 1 - \tau^2/8 \> \> $ (see Eqs. \eqref{omega_N}, \eqref{def xi}).
The extra Maslov phase of $ \> - \pi/2 \> $ acquired when passing the focal point at 
$ \> \xi = 0 \> $, i.e. $ \> \tau \equiv \Omega T = \sqrt{8} \> $ is clearly seen 
\footnote{For better readability phases are plotted in degrees here and in the following figures.} and very accurately
reproduced by Ooura's integration routine.
\newpage
\vspace{1cm}

\begin{figure}[htbp]
\vspace*{-3.8cm}
\bce
\includegraphics[angle=0,scale=0.5]{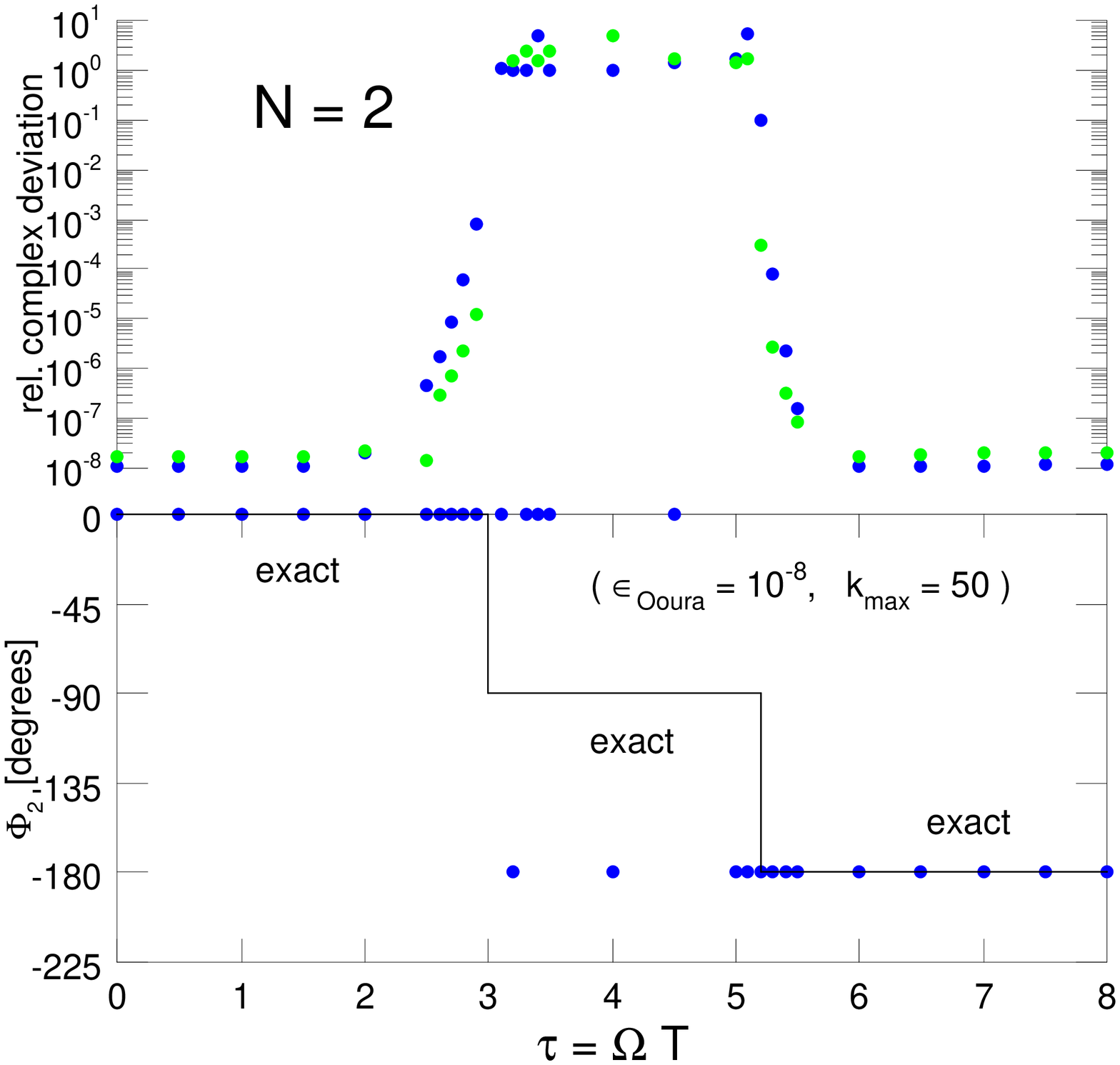}
\label{abb:3}
\ece

{\bf Fig. \arabic{abb}} : {\small Prefactor of the harmonic oscillator propagator as function of the dimensionless time for the 
discretized version of the path integral with $ N = 2 $ (i.e. 3 time slices). 
The lower panel depicts the Maslov phase obtained when passing the two focal points
(solid line: exact, points: numerical result). The upper panel shows the relative complex deviation of the numerical result from the exact one.
The integration over the hyperradius has been performed with
Ooura's rule (with standard accuracy parameters) whereas the integration  over the (only) one angle was done
with the original double exponential (DE) rule of Ref. \cite{TaMori} choosing the same $\epsilon$-parameter as
for the oscillatory integral. Blue or green  points have been obtained with 
$ \> k_{\rm max}^{\rm DE} = 50 \>  {\rm or} \> 500 $. }

\end{figure}

Unfortunately, this does not hold for $ \> N = 2 \> $ as Fig. \ref{abb:3} shows where 
the subsequent integration over the angle $ \phi_1 $ is performed by the double exponential method: 
even by vastly increasing the number of function calls the relative complex deviation from the 
exact result
\be 
F_2^{\rm h.o.}(\tau) \E \sqrt{\frac{m}{8 \pi i \Delta T}} \frac{1}{2 \pi i} \, \int_0^{2 \pi} d\phi_1 \, \int_0^{\infty} dy \, e^{i \omega_2(\phi_1)y}
\E \sqrt{\frac{m}{8 \pi i \Delta T}} \frac{1}{2 \pi} \, \int_0^{2 \pi} d\phi_1 \, \frac{1}{\omega_2(\phi_1) + i 0^+}
\ee
remains large ( {\cal O}(1) ) near the two focal points. This may be due to the distribution
\be 
\frac{1}{\omega_2(\phi_1) + i 0^+} \E \frac{1}{\xi - \sin \phi_1 \cos \phi_1 + i 0^+} \E
{\cal P} \frac{1}{\xi - \sin \phi_1 \cos \phi_1} - i \pi \, \delta \lrp \xi - \sin \phi_1 \cos \phi_1 \rrp
\label{distri}
\ee
which Ooura's routine tries to mimick by a $ 2 k_{\rm max} + 1 $ finite number of terms but the subsequent double exponential integration rule cannot handle properly...

\refstepcounter{abb}

\vspace*{0.2cm}

\begin{figure}[htbp]
\vspace*{-3cm}
\bce
\includegraphics[angle=0,scale=0.4]{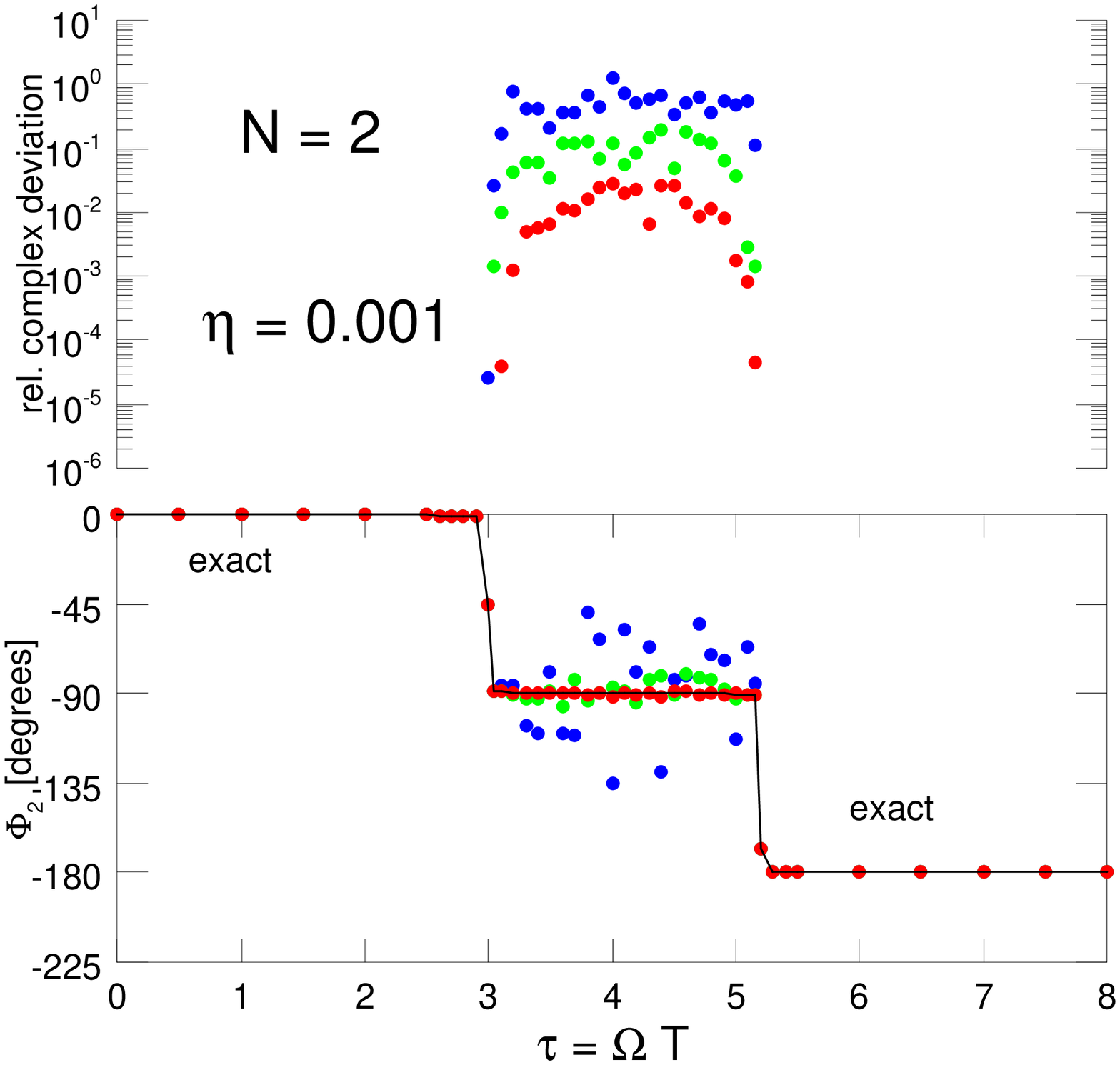}
\label{abb:4}
\ece
\vspace*{0.5cm}

{\bf Fig. \arabic{abb}} : {\small Same as in Fig. 3 but now with an additional damping $ \exp(-\eta y ) $ and $ \eta = 0.001 $. An adaptive
Gauss-Chebyshev rule \cite{PSM} was used for integration over the only one angle with the maximal number of function calls 
$ m_{\rm Cheby} = 1023 $ (blue points), $ 4095 $ (green points), $ 8191 $ (red points).} 
\end{figure}

As an ad-hoc way out of this dilemma one may introduce an overall damping factor
\be 
\boxed{
D_{\eta}(y) \Def e^{- \eta \, y } \> , \quad \eta > 0 \quad {\rm "small"}
\label{damping}
}
\ee
into the $y$-integral. As shown in Fig. \ref{abb:4} this works for 
a damping factor $ \> \eta = 0.001 \> $ both for the standard double exponential integration routine
as well as for a modified adaptive Gauss-Chebyshev integration method. 
\vspace{0.5cm}

\noindent
Indeed,
from Eq. \eqref{FN hyper} it is seen that introducing the {\it ad hoc}-damping factor \eqref{damping} amounts to replacing
\be 
\xi \quad \longrightarrow \quad \xi + i \eta
\ee
or equivalently to shifting the focal points slightly into the complex plane. Then Eq. \eqref{distri} changes into 
\be
\frac{1}{\omega_2(\phi_1) + i 0^+} \> \longrightarrow \>  \frac{1}{\omega_2(\phi_1) + i \eta} \E
\frac{\omega_2(\phi_1) }{\omega_2^2(\phi_1) + \eta^2} - i \frac{\eta}{\omega_2^2(\phi_1) + \eta^2}
\ee
which is finite and a well-known representation of these distributions for $ \> \eta \to 0 \> $.
The exact value of the harmonic oscillator prefactor with damping is worked out in Appendix A, Eqs. \eqref{phase with eta}, \eqref{modulus with eta}.

\refstepcounter{abb}

\refstepcounter{abb}

\begin{figure}[htbp]
\vspace*{-1.5cm}
\bce
\includegraphics[angle=0,scale=0.5]{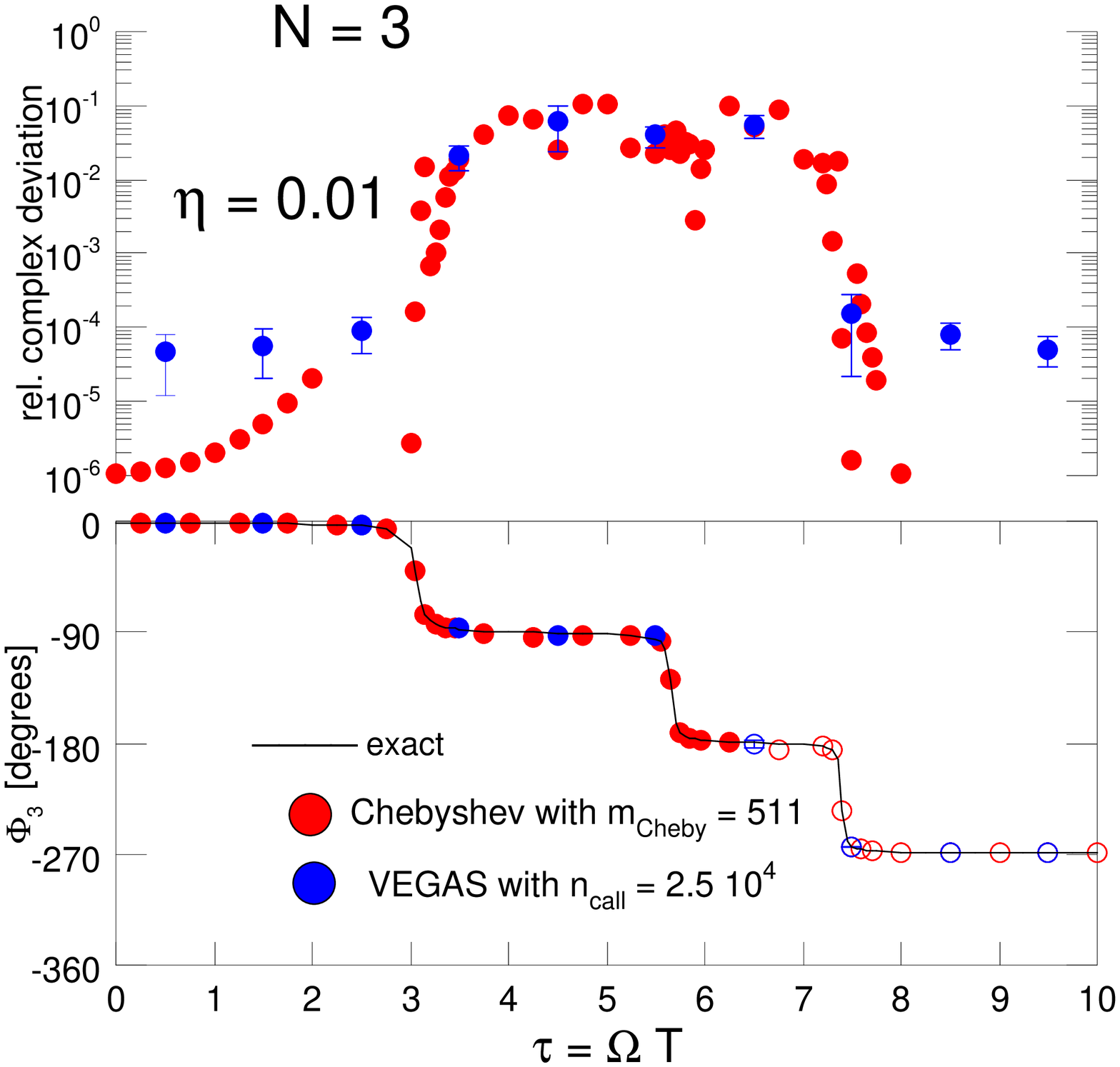}
\label{abb:5}
\ece
\vspace*{0.5cm}

{\bf Fig. \arabic{abb}} : {\small Same as in Fig. 4 but for  $ \> N = 3 $ and damping $\eta = 0.01 $. Results with the adaptive
Gauss-Chebyshev rule for integration over the the two angles angles (maximal number of function calls 
$ m_{\rm Cheby} = 511 $, required relative accuracy $10^{-5}$ )  are compared with the ones 
from the Monte-Carlo routine VEGAS with $n_{\rm call} = 2.5 \cdot 10^4 $ function calls.
Open circles indicate values where the $ n = - 1 $ alignement \eqref{Phi ambiguity} has been applied.}

\end{figure}
\vspace{0.1cm} 

As shown in Fig. 5 the 
adaptive integration routine \cite{PSM} (slightly modified to allow complex integrands and relative complex error to be achieved) and the classic VEGAS program \cite{Vegas} (applied to
real and imaginary parts separately)
give nearly identical results when the same number of function calls is used. The adaptive routine does a little bit better away from the focal points but not in their vicinity. The advantage of the Monte-Carlo 
evaluation is that it also provides  error estimates for the real and imaginary part of the integral which -- by error propagation -- allow an estimate of the error in phase and magnitude of the path integral prefactor. However, due to the delicate integrand these estimates typically are too small by a factor of two and more.
\vspace{0.1cm}

Table 2 displays the numerical results for the ($N=3$)-Gauss-Chebyshev results depicted in Fig. 5. The 
complex harmonic oscillator prefactor starts in the 4th quadrant and moves clockwise with increasing
time. After  passing through 3 focal points the accumulated Maslov phase is $\>  < - \pi \> $. The numerical result then enters the second quadrant which may be interpreted as a positive phase. Choosing $ n = - 1 $ in Eq. \eqref{Phi ambiguity} "alignes" it with the analytic result and allows a meaningful comparison. As  a well-defined Fresnel integral the numerical result for the prefactor -- as given in the second column of Table 2 --
is, of course, unambigous and does not depend on the chosen branch of a multi-valued function as discussed in Ref. \cite{Vivo}.

\clearpage

\bce
\begin{table}[hbt]
\begin{tabular}{r|c|c|r} \hline
                    &                               &                         &       \\
$\tau $ \hspace*{1cm} & \hspace*{2cm}$F_3^{\rm h.o.}/F_3^{\rm free}$ \hspace*{2cm}& \qquad $\delta_{\rm rel}[F_3^{\rm h.o.}]$ \hspace*{1.5cm} &  \hspace*{0.5cm}$\Phi_3^{\rm Maslov}$ 
[degrees] \hspace*{0.5cm} \\
 
                    &                               &                         &    \\ \hline
                    &                               &                         &     \\
     0.0  \hspace*{1cm}     &    $ +0.999  - 0.025 \>  i  $      &  $ 1.05 \cdot 10^{-6} $ &  
     {\footnotesize $(n=0)$} \qquad - 1.4 \quad \\                           
     0.5   \hspace*{1cm}    &    $ +1.019  - 0.026 \> i  $      &  $ 1.22 \cdot 10^{-6} $ &   - 1.5 \quad \\
     1.0   \hspace*{1cm}    &    $ +1.084  - 0.029 \> i  $      &  $ 2.01 \cdot 10^{-6} $ &   - 1.6 \quad \\
     1.5   \hspace*{1cm}    &    $ +1.214  - 0.037 \> i  $      &  $ 4.90 \cdot 10^{-6} $ &   - 1.8 \quad \\
     2.0  \hspace*{1cm}     &    $ +1.464  - 0.057 \> i  $      &  $ 2.03 \cdot 10^{-5} $ &   - 2.2 \quad \\
     2.5  \hspace*{1cm}     &    $ +2.043  - 0.124 \> i  $      &  $ 5.98 \cdot 10^{-14}$ &   - 3.5 \quad \\
     3.0  \hspace*{1cm}     &    $ +5.266  - 2.014 \> i  $      &  $ 2.74 \cdot 10^{-6} $ &   -20.9 \quad \\
     3.5  \hspace*{1cm}     &    $ +0.079 -  2.563 \> i  $      &  $ 1.87 \cdot 10^{-2} $ &   -88.2 \quad \\
     4.0  \hspace*{1cm}     &    $ +0.094 -  2.128 \> i  $     &  $ 7.47 \cdot 10^{-2} $ &   -87.5 \quad \\
     4.5   \hspace*{1cm}    &    $ +0.040 -  1.917 \> i  $     &  $ 2.51 \cdot 10^{-2} $ &   -88.8 \quad \\
     5.0  \hspace*{1cm}     &    $ +0.047  - 2.029 \> i  $      &  $ 1.04 \cdot 10^{-1} $ &   -88.7 \quad \\
     5.5  \hspace*{1cm}     &    $ -0.475  - 4.235 \> i $      &  $ 2.25 \cdot 10^{-2} $ &   -96.4  \quad \\
     6.0  \hspace*{1cm}     &    $ -2.797 - 0.085 \> i  $      &  $ 2.54 \cdot 10^{-2} $ &   -178.2 \quad \\
     6.5  \hspace*{1cm}     &    $ -2.016 - 0.113 \> i  $      &  $ 5.18 \cdot 10^{-2} $ &   -176.8 \quad \\
     7.0  \hspace*{1cm}     &    $ -2.038 + 0.020 \> i $ &  $ 1.94 \cdot 10^{-2} $ & +179.4 $ \> $
     $\stackrel{n=-1}{\longrightarrow}$ \quad -180.6 \quad \\
     7.5  \hspace*{1cm}     &    $-0.316 + 2.935 \> i $  &  $ 1.59 \cdot 10^{-11} $ & +96.2 \quad $\longrightarrow$ \quad -263.8 \quad \\
     8.0  \hspace*{1cm}     &    $ -0.025  + 0.999 \> i $ &  $ 1.05 \cdot 10^{-6} $  & +91.4 \quad $\longrightarrow$ \quad -268.6 \quad \\
     8.5  \hspace*{1cm}     &    $ -0.009  + 0.606 \> i$ &  $ 1.62 \cdot 10^{-8} $ & +90.9 \quad $\longrightarrow$ \quad-269.1 \quad\\
     9.0 \hspace*{1cm}      &    $ -0.005 + 0.421 \> i$ &  $ 7.55 \cdot 10^{-10} $ & +90.7 \quad $\longrightarrow$ \quad -269.3 \quad \\
     9.5  \hspace*{1cm}     &    $ -0.003 + 0.312 \> i$ &   $ 6.53 \cdot 10^{-11} $ &+90.5 \quad $\longrightarrow$ \quad -269.5 \quad \\
     10.0 \hspace*{1cm}     &    $ -0.002 +  0.242 \> i$ &  $ 8.54 \cdot 10^{-12} $ & +90.4 \quad $\longrightarrow$ \quad -269.6 \quad \\
                    &                               &                          &      \\ \hline
\end{tabular}

\vspace{0.8cm}

{\bf Table 2}: 
{\small The (complex) harmonic oscillator prefactor (normalized to the free case) for $ N = 3 $ obtained numerically  with Ooura's and Gauss-Chebyshev integration routines (damping $ \eta = 0.01 $ ) as function of the dimensionless time $ \tau = \Omega T $. With no damping the focal points occur at $ \tau = 3.061, 5.656 , 7.391 $ (see Eq. \eqref{focalN3}) 
where nearly step-like changes of the phase can be seen in Fig. 5).
The third column gives the relative complex deviation of the numerical result from the exact value. 
In the last column the calculated Maslov phase is listed together  with the "alignement"
(addition of $ 2 \pi \, n  , \> \> n = - 1 $ , see Eq. \eqref{Phi ambiguity} )
when the prefactor enters  the second quadrant in the complex plane at $ \tau \simeq 7 $.
}

\end{table}
\ece
For larger values of $ N $ deterministic integration routines become inefficient (the infamous "curse of dimensions") so that only stochastic methods remain. 
Figs. 6 and 7 demonstrate that the VEGAS Monte-Carlo method still works for $ N = 5 $ and 
$ N = 8 $ 
although larger values of the damping parameter $ \eta $ are required to get stable results. 
Longer Monte-Carlo runs (one data point with $10^6$ function calls in Fig. 7 took about 25 minutes on a standard 2.5 GHz PC) are also feasible to improve the statistics.

\refstepcounter{abb}

\begin{figure}[htbp]
\vspace*{-4cm}
\bce
\includegraphics[angle=0,scale=0.5]{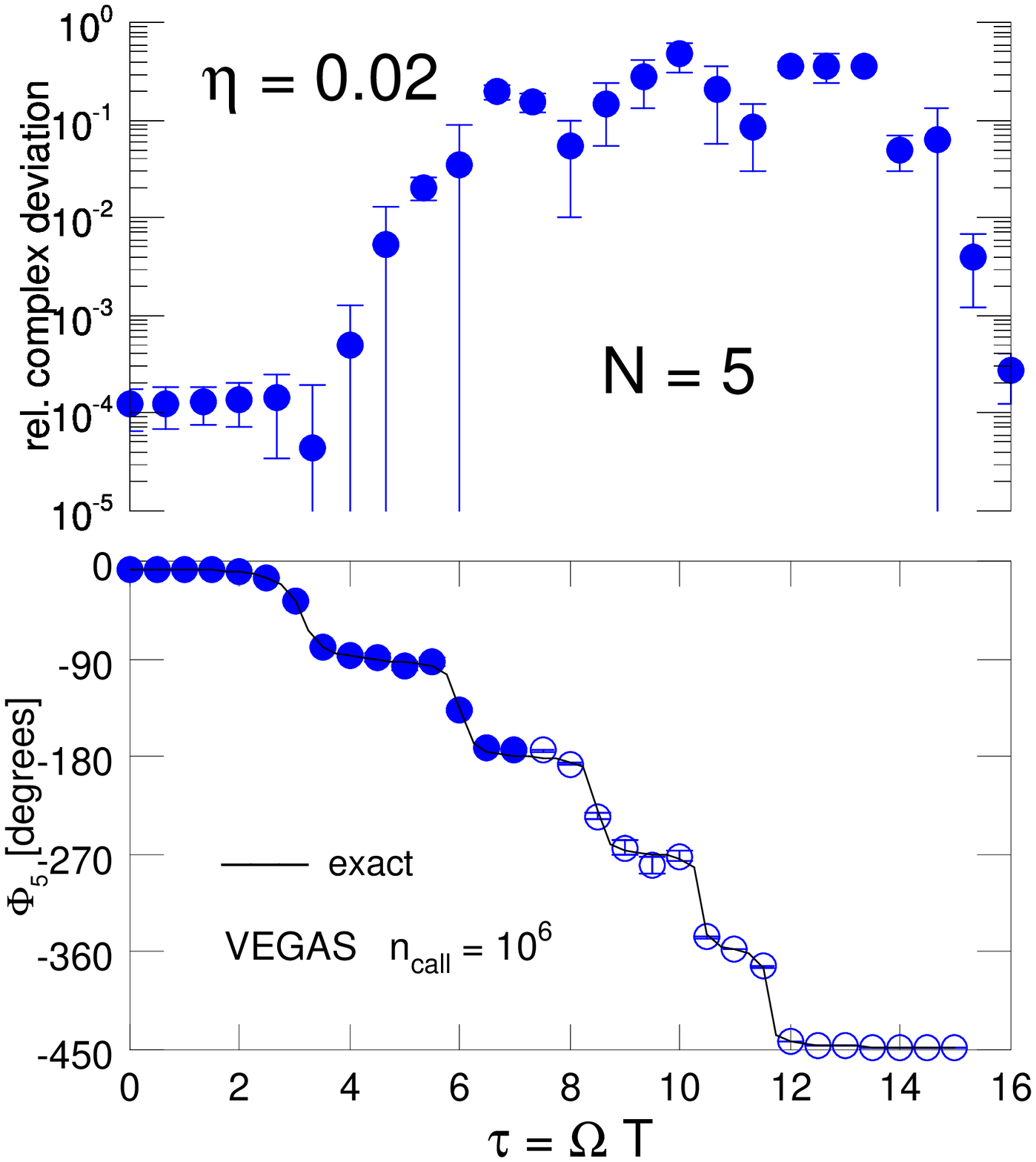}
\label{abb:6}
\ece

{\bf Fig. \arabic{abb}} : {\small Same as in Fig. \ref{abb:5} but for $ \> N = 5 \> $ and  
$ \> \eta = 0.02 \> $.}
\vspace*{1.5cm}

\end{figure}


\refstepcounter{abb}

\begin{figure}[htbp]
\vspace*{-1cm}
\bce
\includegraphics[angle=0,scale=0.5]{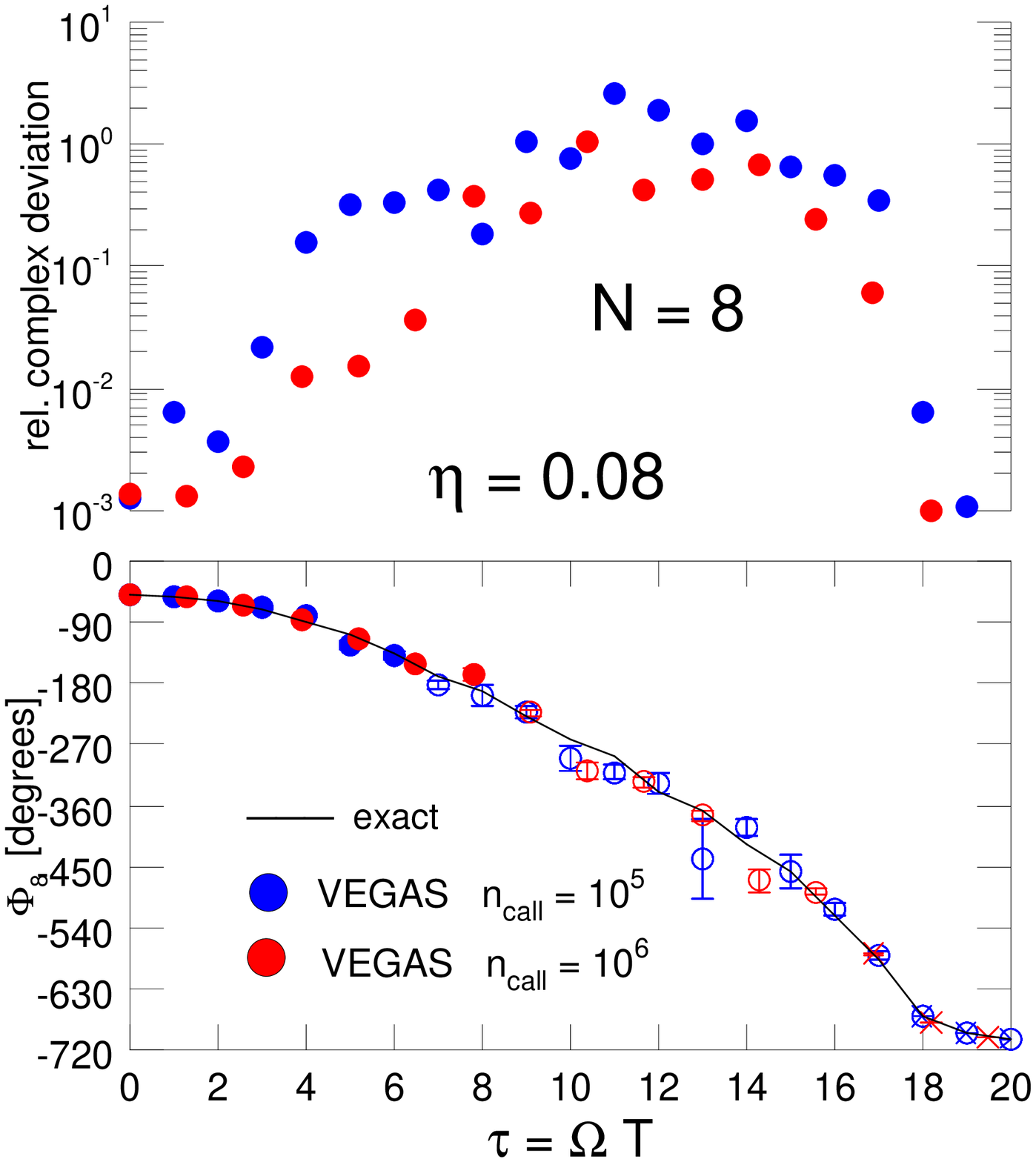}
\label{abb:7}
\ece
\vspace*{0.3cm}

{\bf Fig. \arabic{abb}} : {\small Same as in Fig. \ref{abb:5} but for $ \> N = 8 \> $ and  
$ \> \eta = 0.08 \> $. For the last 3 data points (indicated by crosses) an alignement 
\eqref{Phi ambiguity} with $ n = - 2 $ was applied.}

\end{figure}
\clearpage

\section{Summary and Outlook}
Real-time path integrals for dynamic quantum processes are a numerical challenge as
one has to steer between conflicting requirements: on the one hand
the time-step $ \> \Delta T \> $ has to be small to reach the continuum limit while the dimension $ \> N \> $ of the integrals has to be large enough to capture the relevant time scale $ \> T = (N + 1) \Delta T \> $.
While this can be handled in imaginary time and is widely used to obtain information on static properties
time-dependent processes like scattering require (functional) integration 
over rapidly oscillating functions.
\vspace{0.1cm}

As prototype for these challenges 
I have evaluated numerically Gauss-Fresnel oscillatory integrals of dimension up 
to $ \> N = 20 \> $ , i.e. basically the free particle propagator.
The key to a successfull achievment was Ooura's double exponential integration method \cite{Ooura}
combined with the use of hyperspherical co-ordinates to isolate the most rapidly 
oscillating degree of freedom.
\vspace{0.05cm}

In a second application these tools allowed to calculate numerically 
the prefactor in the harmonic oscillator propagator and thereby the Maslov phase which
emerges each time when the quantum particle passes through a (singular) focal point.
In the discretized version
of the path integral the number of focal points equals the dimension of the oscillatory integral
over hyperradius and angles which in the present work went up to $ \> N = 8 \> $. 

Unfortunately these singularities 
required an additional small damping factor in order to obtain stable results.
Nevertheless this may be considered as an encouraging
step to evaluate real-time path integrals directly e.g for scattering in a finite-range
potential \cite{Ros1}, \cite{Ros2}. This is due to several reasons:  First, the unwanted damping may be dealt with by an extrapolation of the results to zero damping 
similar to the extrapolation to small quark (or pion) mass in lattice gauge
theories \cite{Rothe}. In addition, one may expect that "caustic" singularities 
will be avoided or alleviated in exact path integral formulations for the $T$-matrix 
which go beyond the semiclassical approximation. Also the peculiar properties of the harmonic
potential will be absent in short-range interactions.

While it seems that $ \> N = 8 \> $ is a far cry from the true continuum limit ( $ N \to \infty $ ) of functional integrals improved effective actions \cite{Serbia 1} 
may allow larger time steps and thus fewer time slices. Whether this leads to a reliable
numerical evaluation of functional integrals for scattering requires further investigation.

\vspace{7cm}

\noindent
{\bf Acknowledgement:} I would like to thank Matthias who enabled me to perform the numerical calculations on my home computer and Michael Spira for supplying me with his version of the VEGAS program and for the hospitality in the PSI Particle Theory Group.

\newpage
\bce
{\Large\bf Appendix}
\ece

\renewcommand{\thesection}{\Alph{section}.}
\renewcommand{\theequation}{\thesection\arabic{equation}}
\setcounter{equation}{0}
\setcounter{section}{0}

\vspace{0.2cm}

\section{Maslov phase in the discretized path integral}

Here I give the results for the prefactor in the discretized path integral for the harmonic oscillator, first for a few low-dimensional cases and then for an arbitrary number of time slices.
This then allows to study the continuum limit.

\vspace{0.2cm}

\noindent
\underline{\bf $ N = 1 $}:

\be
F_1^{\rm h.o.}(T) \E \frac{m}{2 \pi i \Delta T }  \, \Uint dx_1 \> 
\exp \lcp \frac{i m}{2 \Delta T} \lrp 2 - (\Delta T)^2 \Omega^2 \rrp x_1^2 \rcp 
\E \lrp \frac{m}{2 \pi i \Delta T} \rrp^{1/2} \, \frac{1}{\sqrt{2 \xi}} 
\ee
where  $ \> \xi \> $ has been defined in Eq. \eqref{def xi}. It is seen that in this rough approximation for the path integral (just 2 time slices!) a focal point occurs at $ \xi_1 = 0 $ where the argument of the inverse square root turns negative. 
Using $ \Delta T = T/2 $ this translates into 
\be
\Omega T \E \sqrt 8 = 2.828...  \deF X_1^{(1)}
\ee
compared to the exact continuum value of $ \pi = 3.141..$. Thus
\be 
\Omega \, f_1(\tau) \E \tau \lrp  1 - \frac{\tau^2}{8} \rrp \> , \quad \Phi_1^{\rm Maslov}(\tau) \E 
- \frac{\pi}{2} \Theta \lrp \tau - X_1^{(1)} \rrp \> .
\ee
\vspace{0.2cm}

\noindent
\underline{\bf $ N = 2 $}:
\bea
F_2^{\rm h.o.}(T) \EA \lrp \frac{m}{2 \pi i \Delta T } \rrp^{3/2} \, \Uint dx_1 \, dx_2 \> 
\exp \lcp \frac{i m}{2 \Delta T} \lsp  x_1^2 + x_2^2 + \lrp x_2 - x_1 \rrp^2 - \Omega^2 
(\Delta T)^2 \lrp x_1^2 + x_2^2 \rrp \rsp \rcp \non
\EA \lrp \frac{m}{2 \pi i \Delta T} \rrp^{3/2} \, \Uint dx_1 \, dx_2
\exp \lcp \frac{i m}{\Delta T} \lsp  \xi \lrp \underbrace{x_1 - x_2/2 \xi}_{\deF x_1'} \rrp^2 +
\lrp \xi - \frac{1}{4 \xi} \rrp x_2^2  \rsp \rcp \> .
\eea
After shifting the integration variable $ \> x_1  \to x_1' \> $ one obtains
\be 
F_2^{\rm h.o.}(T) \E \lrp \frac{m}{2 \pi i \Delta T} \rrp^{1/2} \, \frac{1}{\sqrt{\xi} }
\frac{1}{\sqrt{4 \xi - 1/\xi}} \> .
\label{F2}
\ee
With $ \Delta T = T/3 $ one sees that in this approximation there are 2 focal points during the evolution of the harmonically bound particle: with increasing time
one at 
\be 
\xi_1 \E \frac{1}{2} \> \Longrightarrow  \> \Omega T = 3 \deF X_1^{(2)}
\ee
(instead of the exact value $\pi $) and another one at 
\be 
\xi_2 \E  -\frac{1}{2} \> \Longrightarrow \>  \Omega T = 3 \sqrt 3 = 5.196... \deF X_3^{(2)}
\ee
(instead of the exact value $ 2 \pi = 6.283...$). Note that 
\be 
\xi \E 0 \> \Longrightarrow \>  \Omega T = 3 \sqrt 2 = 4. 242... \deF X_2^{(2)}
\ee
is {\it not} a focal point as the prefactor does not diverge here.

\noindent
Thus from Eq. \eqref{F2} one has
\be 
\Omega f_2(\tau) \E \tau \lrp 1 - \frac{\tau^2}{9} \rrp \, \lrp 1 - \frac{\tau^2}{27} \rrp \> .
\ee
Obviously the function $ 4 \xi - 1/\xi $ is negative for $ 0 < \xi < 1/2 $ and for $ \xi < -1/2 $.
With $ \xi = 1 - \tau^2/18 \>  $  the $(N = 2)$- approximation to the path integral thus gives the following Maslov phase
\bea 
\Phi_2^{\rm Maslov}(\tau) \EA - \frac{\pi}{2} \, \lsp \Theta \lrp \tau - X_1^{(2)}  \rrp \Theta \lrp 
X_2^{(2)} - \tau \rrp + \Theta \lrp \tau - X_2^{(2)} \rrp
+ \Theta \lrp \tau - 3 \sqrt 3  \rrp \rsp \non
\EA - \frac{\pi}{2} \, \sum_{n=1, n \neq 2}^3 \Theta \lrp \tau - X_n^{(2)}  \rrp 
\label{Maslov 2}  
\eea 
as function  of $ \> \tau = \Omega T $. Note that naively combining the square roots in Eq. \eqref{F2}
into $ (4 \xi^2 - 1 )^{-1/2} $ (without considering their phases in the complex plane) would give the wrong result that the Maslov phase $ \Phi_2 $ would vanish for $ \xi < - 1/2 $, i.e. 
$ \tau > 3 \sqrt 3 $. Note also that in the final result \eqref{Maslov 2} the point 
$ \tau = X_2^{(2)} $
does not appear: only the true focal points {\it add} a phase $ - \pi/2 $ when the particle passes through it during its time evolution.
\vspace{0.2cm}

\noindent
\underline{\bf $ N = 3 $}:
\bea
F_3^{\rm h.o.}(T) \!&=& \! \lrp \frac{m}{2 \pi i \Delta T } \rrp^2  \! \Uint \! \! dx_1 dx_2  dx_3
\exp \Bigg \{ \frac{i m}{2 \Delta T} \Big [ x_1^2+x_3^2+\big ( x_2 - x_1 \big)^2 + 
\big( x_3 - x_2 \big)^2
- ( \Omega \Delta T)^2 \big ( x_1^2+x_2^2+x_3^2 \big ) \Big ] \Bigg \} \non
\! \EA  \! \lrp \frac{m}{2 \pi i \Delta T } \rrp^2 \! \Uint \! \! dx_1 dx_2 dx_3
\exp \Bigg \{ \frac{i m}{2 \Delta T} \Big [  2 \xi \lrp \underbrace{x_1 - x_2/2 \xi}_{=x_1'} \rrp^2 + \lrp 2 \xi - \frac{1}{\xi} \rrp x_2^2  + 2 \xi \lrp \underbrace{x_3 - x_2/2 \xi}_{\deF x_3'} \rrp^2  \Big ] \Bigg \} \non
\EA  \lrp \frac{m}{2 \pi i \Delta T} \rrp^{1/2} \> \frac{1}{\sqrt{2 \xi}} \, \frac{1}{\sqrt{2 \xi - 1/\xi}} \, \frac{1}{\sqrt{2 \xi}}  \> .
\eea
With $ \Delta T = T/4 $ one finds that there are 3 focal points at 
\bea
\xi_1 \E \frac{1}{\sqrt{2}} && \Longrightarrow \> \Omega T = 4 \sqrt{2 -\sqrt 2} = 3.0614... \deF X_1^{(3)} \non 
 \xi_2 \E 0 && \Longrightarrow \> \Omega T = 4 \sqrt 2 = 5.6556... \deF X_2^{(3)} \quad \mbox{(here
a genuine focal point with multiplicity 2)} \non
\xi_3 \E - \frac{1}{\sqrt{2}} && \Longrightarrow \>  \Omega T =  4 \sqrt{2 + \sqrt 2} = 7.3910... \deF X_3^{(3)}
\label{focalN3}
\eea
instead of $ \Omega T = n \pi , n = 1, 2, 3 \ldots $ 
and therefore
\bea 
\Omega f_3(\tau) \EA \tau \, \lrp 1 - \frac{\tau^2}{32} \rrp \, \lrp 1 - \frac{\tau^2}{8} + \frac{\tau^4}{512} \rrp 
\\   
\Phi_3^{\rm Maslov}(\tau) \EA - \frac{\pi}{2} \, \lsp \Theta \lrp \tau - X_1^{(3)} \rrp \, \Theta \lrp  X_2^{(3)}  - \tau \rrp
+ 2 \, \Theta  \lrp  \tau - X_2^{(3)} \rrp  + \Theta \lrp \tau - X_3^{(3)}  \rrp \rsp \non
\EA - \frac{\pi}{2} \sum_{n=1}^3 \Theta  \lrp  \tau - X_n^{(3)} \rrp
\eea
\vspace{0.2cm}

\noindent
\underline{\bf $ N $ arbitrary}:
\vspace{0.1cm}

\noindent
The standard method to calculate the prefactor for quadratic Lagrangians in the continuum limit
is the Gel'fand-Yaglom method (see, e.g. Ref. \cite{Schul}, ch. 6 or Ref. \cite{engpath}
ch. 1.3) which leads to a differential equation for $ f(T) $. The same method can also be used  to evaluate $ f_N(\tau) $ for finite $ \Delta T $ by a recurrence relation:
Define $ p_0(\xi) = 1 \> , p_1(\xi) = 2 \, \xi \> $ and calculate
\be 
p_{j+1}(\xi)  \E 2 \xi \> p_j(\xi) - p_{j-1}(\xi) \> , \quad j = 1, 2 \ldots N-1  \> .
\label{recurrence}
\ee
Then
\be 
\Omega f_N(\tau) \E \frac{\tau}{N+1} \, p_N  \lrp \xi = 1 - \frac{\tau^2}{2 (N+1)^2} \rrp \> .
\label{f_Ntau}
\ee
%
%
This recurrence relation may be solved either by standard methods (see, e.g. Ref. \cite{WikiRecurrence}) or by examination of the recurrence relations for the classical orthogonal
polynomials \footnote{See, e.g. Ref. \cite{Handbook}, eq. 22.7.4 for the Chebyshev polynomials of the first kind $ T_N(\xi) $ and eq. 22.7.5 for the ones of the second kind
$ U_N(\xi) $.}. Although both $ T_N(\xi) $ and $ U_N(\xi) $ obey the same recurrence relation,
only the Chebyshev polynomials of the second kind fulfill the initial condition  $ p_1(\xi) = 2 \xi $ and therefore the final result is
\be
p_N(\xi) \E U_N(\xi) \quad \Longrightarrow  \Omega \, f_N(\tau) \E \frac{\tau}{N+1} \, U_N  \lrp 1 - \frac{\tau^2}{2 (N+1)^2} \rrp \> .
\label{pN}
\ee
With $ \> \>  U_0(\xi) = 1, \> 
U_1(\xi) = 2 \xi,  \> U_2(\xi) = 4 \xi^2 - 1, \> U_3(\xi) = 8 \xi^3 - 4 \xi \> \> $  
( Ref. \cite{GradRyz} eq. 8.943) the explicit results obtained above are reproduced.
Since (Ref. \cite{Handbook}, eq. 22. 3.16)
\be 
U_N(\xi) \E \frac{\sin \lrp (N+1) \arccos \xi \rrp}{\sin \lrp \arccos \xi \rrp}
\ee
the zeroes of this function are real and given by the simple expression (Ref. \cite{Handbook},  eq. 22.16.5)
\be 
\xi_n^{(N)} \E \cos \lrp \frac{n}{N+1} \pi \rrp \quad n = 1, 2 \ldots N \> .
\label{zeroes}
\ee
Using the definition \eqref{def xi} this translates into the following formula for the (positive) 
time to reach the $n^{\rm th}$ focal point
\be 
\Omega T_n \equiv X_n^{(N)} \E 2 ( N + 1 ) \, \sin \lrp \frac{n}{2 (N+1)} \pi \rrp
\ee
in which the correct continuum limit $ N \to \infty $ is evident. 
Since the Chebyshev polynomial of the second kind $ U_N(\xi) $ is a polynomial of order $ N $
with real zeros $ \xi_n^{(N)} $ it may be written as
\be 
U_N(\xi) \E C_N \, \prod_{n=1}^N \lrp \xi - \xi_n^{(N)} \rrp
\ee
where the normalization factor $ C_N = 2^N $  is a consequence of the recurrence relation
 \eqref{recurrence} which just gives this factor for the leading power of $ \xi \> $
(alternatively one can employ the explicit expression  Eq. 22.3.7 in Ref. \cite{Handbook}).
Upon taking the inverse square root of $ p_N(\xi) $ for the prefactor (see Eqs. (\ref{prefactor square root}, \ref{f_Ntau}))
each factor contributes a phase of $ - \pi/2 $ whenever it becomes negative, i.e. 
$ \xi \, < \xi_n^{(N)} $.
Thus
\be 
\Phi_N^{\rm Maslov}(\xi) \E - \frac{\pi}{2} \, \sum_{n=1}^N \Theta \lrp \xi_n^{(N)} - \xi \rrp \> ,
\ee
a discontinous function of the time.

The continuum limit for the function $ f_N(\tau) $ in Eq. \eqref{pN} is more involved but may be 
derived by expressing it as a hypergeometric function 
\be 
\Omega f_N(\tau) \E \tau \> \cdot \> _2F_1(a,b;c;z) \E \tau \, \cdot \sum_{n=0}^{\infty} \frac{(a)_n (b)_n}{(c)_n} \, \frac{z^n}{n!} 
\ee 
with
\be
a \E - N \> , \quad b \E N + 2 \>, \quad
c \E\frac{3}{2} \>, \quad  z \E \frac{\tau^2}{4 (N+1)^2} 
\ee
by means of eq. 22.5.48 in Ref. \cite{Handbook}.
Thus
\be
\Omega f_N(\tau) \deF  \sum_{n=0}^N d_n \, \tau^{2n+1} \E \tau  - 
\frac{N^2+2N}{(N + 1)^2} \, \frac{\tau^3}{6} + \frac{(N-1)N(N+2)(N+3)}{(N+1)^4} \, \frac{\tau^5}{120} + \ldots \> .
\ee
The first three expansion coefficients agree with the explicit calculations for $ N = 1, 2, 3 $ and tend to the first three coefficients in the series expansion of $ \> \sin \tau \> $. 
From the series expansion of the hypergeometric function one deduces that the $n^{\rm th} $ term reads
\be 
d_n \E \frac{ (-N)_n (N+2)_n}{ (3/2)_n} \, \frac{1}{n!} \lsp \frac{1}{4 (N+1)^2} \rsp^n \> .
\ee
This can be evaluated by expressing the Pochhammer symbols as factorials 
\be 
(-N)_n \E (-)^n \, \frac{N!}{(N-n)!} \> , \quad (N+2)_n \E \frac{(N+n+1)!}{(N+1)!} \> , \quad
\lrp \frac{3}{2} \rrp_n \E \frac{(2 n + 1)!}{2^{2n} n!} 
\ee
and gives
\be 
d_n \E \frac{(-)^n}{(2n + 1)!} \, \prod_{k=1}^{2n + 1} \lrp \frac{N-n+k}{N+1} \rrp \> 
\stackrel{N \to \infty, \> n \> {\rm fixed}}{\longrightarrow} \> \frac{(-)^n}{(2n + 1)!} \quad \Longrightarrow \quad
\Omega f_N(\tau) \> \stackrel{N \to \infty}{\longrightarrow} \> \sin \tau \> .
\ee

\vspace{0.4cm}

\noindent
\underline{\bf with damping}:                      
\vspace{0.1cm}

\noindent
If the damping factor \eqref{damping} is used to regulate the discrete path integral the variable $ \xi $ is replaced by
\be 
\xi \longrightarrow \bar \xi \Def \xi + i \eta
\ee
which leads to the following modifications:  Knowing that $ p_N(\xi) $ is a polynomial of degree $ N $ with 
zeroes at $ \xi_n^{(N)} $ (see Eq. \eqref{zeroes}) one can immediately write
\be 
p_N(\bar \xi) \E 2^N \, \prod_{n=1}^N \lrp \bar \xi - \xi_n^{(N)} \rrp \E 
2^N \, \prod_{n=1}^N \lrp \xi -\xi_n^{(N)} + i \eta \rrp 
\ee
As before this product formula can now be used to calculate the prefactor of the
damped harmonic oscillator path integral with each factor contributing separately to 
the inverse (principal value) square root
\be 
F_N^{({\rm damped} \> {\rm h.o)}}(T;\eta) \E \lrp \frac{m}{2 \pi i \Delta T} \rrp^{1/2} \, 2^{-N/2} \, \prod_{n=1}^N  \Bigl [ \xi - \xi_n^{(N)} +  i \eta \Bigr ]^{-1/2} \> , \quad \xi \E 1 - 
\frac{1}{2} \lrp \frac{\Omega T}{N+1} \rrp^2 \> .
\ee
From this one can read off the exact Maslov phase as
\be 
\Phi_N^{\rm Maslov, damped}(\tau;\eta) \E - \frac{1}{2} \sum_{n=1}^N \, {\rm arg} \lrp \xi - \xi_n^{(N)} + i \eta \rrp 
\label{phase with eta}
\ee
with the standard definitions for the argument of a complex quantity
\begin{subequations} 
\bea 
{\rm arg}\, (z = x + i y) \EA \arctan \frac{y}{x} + \pi \, \Theta(-x) \, \sgn \, y \> , 
\quad - \pi < {\rm arg} \, z \le \pi \> , \> \>    -\frac{\pi}{2} \le \arctan \frac{y}{x} \le 
\frac{\pi}{2} \\
&\Eqi & {\rm atan}2(y,x) 
\label{atan2}
\eea
\end{subequations}
Note that the phase is now a continous function of $ \xi $, i.e. of time $\tau = \Omega T $.
The modulus of the prefactor is
\be 
\Bigl | F_N^{({\rm damped} \> {\rm h.o)}}(\xi;\eta) \Bigr | \E \lrp \frac{m}{2 \pi \Delta T} \rrp^{1/2} \, 
2^{-N/2} \prod_{n=1}^N  \lsp \lrp \xi - \xi_n^{(N)}\rrp^2 +   \eta^2 \rsp^{-1/4}\> .
\label{modulus with eta}
\ee

\vspace{1cm}

\section{Some numerical details: Step size and accuracy}
\setcounter{equation}{0}

Inevitably the sum over $k$ in Eq. \eqref{Ooura form2} has to  be restricted to a finite 
number of terms: $ \> |k|  \le k_{\rm max} \> $. Then we need to know which step size $ h $
one has to take to get results accurate up to a given precision. This can be determined by examining 
the asymptotic behaviour of $ \phi(t= k h) \> $. Let us write
\be
\phi(t) \E t + t \frac{e^{-a(t)}}{1 - e^{-a(t)}} \E t + t \frac{1}{e^{a(t)} - 1} \quad {\rm with} \quad a(t) \Def 2 t + \alpha \lrp 1 - e^{-t} \rrp + \beta \lrp e^t - 1 \rrp 
\ee
and consider first the limit $ \> t \to + \infty $.
Obviously
\be 
a(t)  \> \stackrel{t \to +\infty}{\longrightarrow} \> \beta e^t + 2 t + \alpha - \beta + {\cal O} 
\lrp e^{-t} \rrp
\ee
grows exponentially
and the factor in the curly brackets in Eq. \eqref{Ooura form2} becomes
\bea 
\Big \{ \exp \lsp i \, {\rm sgn}(\omega)\, \frac{\pi}{h} \, 
\phi(k h)\rsp - (-1)^k \Big \} \EA (-1)^k \, \lcp \exp \lsp i \, {\rm sgn}(\omega) \, 
\frac{k \pi}{e^{a(k h)} - 1} \rsp - 1 \rcp \non
&  \stackrel{k h  \to +\infty}{\longrightarrow}  & (-1)^k \> \lcp i \, {\rm sgn}(\omega) \, 
k \pi \, e^{-a(k h)}+ {\cal O} \lrp e^{-2a(k h)} \rrp \rcp
\label{small terms}
\eea
leading to a double exponential decay of terms with $ k \simeq k_{\rm max} $ as desired
\vspace{0.3cm}

Requiring a suppression to size $ \epsilon_{\rm Ooura} $ we demand (cf. eq. (5. 19) in Ref. \cite{num osc})
\be 
\exp \lsp - \beta e^{k_{\rm max} h} \rsp \le \epsilon_{\rm Ooura} \quad \Longrightarrow
h \> \ge \>  \frac{1}{k_{\rm max}} \, \ln \lrp - \frac{1}{\beta} \ln \epsilon_{\rm Ooura} \rrp \>.
\label{eps beta}
\ee
Since the step size $ h $ should be as small as possible this translates into an equality, i.e. 
a determination of the 
step size once the maximal number $ \> 2 k_{\rm max} + 1 \> $ of function calls and the accuracy parameter $ \epsilon_{\rm Ooura} $ are chosen.
For large negative values of $ t = k h $ one finds 
\be 
\phi(t)  \stackrel{t \to -\infty}{\longrightarrow}  - \exp \lrp  - \alpha e^{|t|} + 2 t 
+ {\cal O}(\ln |t|)\rrp \> \Longrightarrow \> \phi'(t) \> \longrightarrow \> - \alpha \, \exp \lrp - \alpha e^{|t|} + {\cal O}(t, \ln |t|) \rrp \> .
\ee 
Since $ \phi'(k h ) $ is the weight in Ooura's integration rule the terms
with $ \> k \simeq -k_{\rm max} \> $ will be suppressed double exponentially if
\be 
\exp \lsp - \alpha e^{k_{\rm max} h} \rsp \le \epsilon_{\rm Ooura} \quad \Longrightarrow
h \ge  \frac{1}{k_{\rm max}} \, \ln \lrp - \frac{1}{\alpha} \ln \epsilon_{\rm Ooura} \rrp \>.
\label{eps alpha}
\ee
As $ \alpha < \beta $ this is a stronger requirement than the one in Eq. \eqref{eps beta}. Note
that Eq. \eqref{eps alpha} is an implicit equation for the step size as $ \alpha = \alpha(h) \> $
(see Eq. \eqref{alpha beta}). We solve it by a few iterations starting with 
$ \alpha_{(0)} = \beta $ .

\vspace{0.2cm}

Calculations in this work have been performed in double-precision arithmetic.
Nevertheless the numerical implementation require some care when evaluating
small terms as 
"smallness by subtraction" leads to a loss of accuracy. Here this is exacerbated by the fact that these terms are weighted by a (large) power of the hyperradius
which peaks at $ k \simeq k_{\rm max} $ although finally made small by the double exponential 
decay. One could use a power series expansion as indicated in the last line of Eq. \eqref{small terms} but a simpler remedy is not to evaluate directly $ \> \exp(ix) - 1  \> $
for small $ x $ but instead  $ \> \exp(ix) - 1 \E 2 i \, \sin(x/2) \, \exp(ix/2) \> $, 
i.e. making them "small by multiplication".

\end{document}